\documentclass{jfm}
\usepackage{graphicx}
\usepackage{subfig}
\usepackage{mathtools}
\usepackage{xcolor}
\usepackage{numprint}
\usepackage{braket}
\usepackage{tikz}
\usetikzlibrary{plotmarks}
\newcommand\marksymbol[2]{\tikz[#2,scale=1.2]\pgfuseplotmark{#1};}
\usepackage{multirow,stackengine}
\begin{document}
\newcommand{\red}[1]{\textcolor{black}{#1}}

\newcommand{\blueline}{\raisebox{2pt}{\tikz{\draw[-,blue,dashed,line width = 1.0pt](0,0) -- (10mm,0);}}}

\newcommand{\blackline}{\raisebox{2pt}{\tikz{\draw[-,black,dashed,line width = 0.5pt](0,0) -- (10mm,0);}}}

\shorttitle{Separation scaling for  reconnection} 
\shortauthor{Jie Yao and Fazle Hussain} 

\title{Separation scaling for viscous vortex reconnection}

\author{
 Jie Yao\aff{1},
  \and 
 Fazle Hussain\aff{1}
   \corresp{\email{fazle.hussain@ttu.edu}},
 }

\affiliation{
\aff{1}
Department of Mechanical Engineering, Texas Tech University,  Lubbock, Texas, USA, 79409
}

\maketitle

\begin{abstract}
Reconnection plays a significant role in the dynamics of plasmas, polymers and macromolecules, as well as in numerous laminar and turbulent flow phenomena in both classical and quantum fluids.
Extensive  studies in quantum vortex reconnection show that the minimum separation distance $\delta$ between  interacting vortices  follows a $\delta\sim t^{1/2}$  scaling. 
Due to the complex nature of the dynamics (e.g., the formation of bridges and threads as well as successive reconnections and avalanche), such scaling has never been reported for  (classical) viscous vortex reconnection.
Using  direct numerical simulation of the Navier-Stokes equations, 
we study viscous reconnection of slender vortices, whose  core size is much smaller than the radius of the vortex curvature. 
For separations that are large compared to the vortex core size, 
we discover that $\delta(t)$ between the two interacting viscous vortices surprisingly also follows the  1/2-power scaling for both pre- and post-reconnection events.
The prefactors in this 1/2-power law  are found to depend not only on the initial configuration but also on the vortex Reynolds number (or viscosity).
Our finding in viscous reconnection, complementing numerous works on quantum vortex reconnection,  suggests that there is indeed a universal route for reconnection -- an essential result for understanding the various facets of the  vortex reconnection phenomena and their  potential modeling, as well as possibly explaining turbulence cascade physics.

\end{abstract}

\section{Introduction}

Reconnection,  a fundamental topology-transforming event,  has been a subject of  intense recent study  in both classical \citep{kida1994vortex,pumir1987numerical,melander1989cross,kleckner2013creation} and quantum  \citep{koplik1993vortex,barenghi2001quantized,bewley2008characterization,paoletti2010reconnection} fluids, as well as in many other fields, such as  plasmas \citep{priest_forbes_2000}, polymers, and macromolecules \citep{vazquez2004tangle}.
In turbulent flows, vortex reconnection  appears to be the main mechanism for  energy cascade: i) in quantum fluids, reconnection excites a cascade of Kelvin waves leading to energy dissipation via emissions of phonons and rotons \citep{kivotides2001kelvin,vinen2003kelvin}; 
ii) in classical fluids, finer and finer scales and turbulence avalanche can occur through successive reconnections \citep{melander1989cross,yao2020physical}. Reconnection is also believed to  play an essential  role in several other physical phenomena, such as  fine-scale mixing \citep{hussain1986coherent}, and  noise generation \citep{leadbeater2001sound}.

One simple but important question in reconnection is the time scaling  of the minimum distance $\delta(t)$ between the two  interacting vortices. Assuming that the reconnection is a local process in space and the circulation $\Gamma$ is the only relevant dimensional quantity involved,  dimensional analysis yields 
\begin{eqnarray}\label{eqn:dsca}
\delta(t)=A^{\pm}(\Gamma |t-t_0|)^{1/2},
\end{eqnarray}
where $t_0$ is the reconnection time, and $A^{-}$ and $A^{+}$ are dimensionless factors for pre- and post-reconnection, respectively.
Such a 1/2-power scaling has been numerically observed for reconnection of line vortices using the Biot-Savart  law \citep{PhysRevLett.72.482, kimura2017scaling} and also for reconnection of quantized vortices by  integrating the Gross-Pitaevksii  equation
\citep{nazarenko2003analytical,villois2017universal}.
In addition, recent quantum experiments \citep{paoletti2010reconnection,fonda2019reconnection}  confirmed  this scaling when the  distances between two interacting vortices  are large compared with the  vortex diameter but small compared with those from other adjacent vortices. 
Note that deviations from this 1/2 scaling were also reported in several works \citep{zuccher2012quantum,allen2014vortex,rorai2016approach}. 

\begin{figure}
\centering
\includegraphics[width=.7\linewidth]{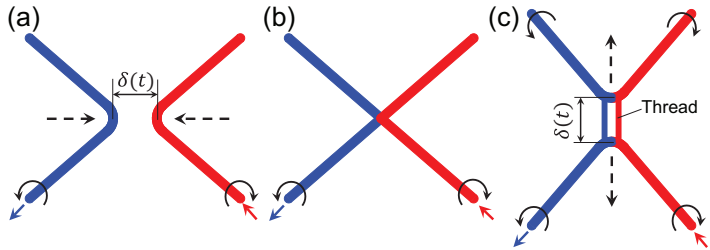}
\caption{Schematic of the evolution of (classical) viscous  vortex reconnection: (a) before; (b) during, and (c) after reconnection. The  curved arrows indicate the rotating directions of the vortices; and the dashed straight arrows represent the directions of vortex motion. Note that the actual reconnection, which is intrinsically three-dimensional, is never complete in classical fluids, leaving  unreconnected  parts  as threads. }
\label{fig:VR_INIT}
\end{figure}

In contrast to the vast literature on the time scaling of $\delta(t)$  in quantum fluids,  very limited results have been reported for reconnection in classical fluids, which are  governed by the Navier-Stokes (N-S) equations (figure \ref{fig:VR_INIT}).
By performing the direct numerical simulation (DNS) of two antiparallel vortex tubes reconnection,  \cite{hussain2011mechanics} 
found  that the minimum distance $\delta$ between the vortex centroids scales asymmetrically as $(t_0-t)^{3/4}$ and $(t-t_0)^{2}$  before and after the reconnection. 
Note that in this study, the vortex core size $\sigma$ is comparable to the initial separation distance $\delta$ between these vortices (i.e., $\sigma/\delta\approx 0.4$) -- which definitely breaks the local assumption required for the $1/2$ scaling.
Inspired  by the recent works of    \cite{moffatt2019towards1,moffatt2019towards2}  on the finite time singularity of  Euler and  Navier-Stokes equations, we studied reconnection of  two colliding slender vortex rings (the ratio between the initial vortex core size $\sigma$ and the  radius of the ring $R$ is approximately 0.01)  and found  that $\delta(t)$ before reconnection follows a $1/2$ scaling when $\sigma\ll \delta \ll R$ \citep{yao2020singularity}.
 The main objective of  the present work is to further elucidate the time scaling of minimum separation distance  for (classical) viscous vortex reconnection.  
In particular, we want to address the following questions:
i)  does the time scaling of the minimum distance follows  $\delta\sim t^{1/2}$ scaling for both before and after reconnection?
ii) what dictates  the prefactors in the scaling?
and 
iii) what are the  similarities/differences between classical and quantum vortex reconnections?  

\section{Results}

Previous studies of the dynamics of slender vortices are mainly based on  the vortex filament (VF) method, which is based on the B-S law \citep{siggia1985collapse,PhysRevLett.72.482,kimura2018tent}. 
To regularize the singular kernel of the B-S integral, a cutoff needs to be employed. 
With such regularization, the B-S integration always diverges near the singular time of reconnection \citep{villois2017universal,kimura2018tent}.
An ad hoc ``cut-and-paste'' algorithm is typically required for studying post-reconnection scenario \citep{schwarz1985three,baggaley2012sensitivity,Galantucci12204}. 
However, as reconnection in classical fluids is very complex, such an algorithm is very difficult to  implement. 
Hence, the VF method is mainly employed for studying  the pre-reconnection event. 

With the rapid development of supercomputers these days,  DNSs for considerably large-scale flow problems are becoming feasible.
Here, we aim to employ DNS of the N-S equations for studying viscous reconnection of  slender vortices.
The numerical method employed here is the same as those used in \cite{yao2020physical}. 
To understand what is universal in reconnections, three different  vortex configurations are considered. 
Case I is two colliding vortex rings, which is the same as that   in \cite{moffatt2019towards1, moffatt2019towards2,yao2020singularity} for studying the possible formation of finite time singularity of Euler and N-S equations.  
Case II  is the  two initially rectilinear, orthogonal vortices, which corresponds to the limit where the radius of curvature $\kappa$ of two vortices are extremely large. 
Finally, to study the interaction of vortices with significantly different curvatures, following \cite{Galantucci12204}, we also consider a case of a vortex ring interacting with an isolated vortex tube (Case III). 
For all cases, the initial vorticity distribution in the cross-section is assumed to be Gaussian $\omega(r)=\Gamma_0/(4\pi \sigma^2_0)\exp[-r^2/4\sigma^2_0]$ with the circulation $\Gamma_0=1$ and core scale $\sigma_0=0.01$.
\textcolor{black}{Compared with those in the past  studies \citep{kida1994vortex,boratav1992reconnection,melander1989cross,chatelain2003reconnection}, the distinction of our simulations is the  larger ratio of the radius of curvature to the core size (i.e., $R_0/\sigma_0\ge100$). }
 As the viscous effect is an essential issue in classical fluids,  for each configuration,   two different Reynolds numbers ($Re_\Gamma\equiv\Gamma_0/\nu=2000$ and $4000$),  achieved by changing  the kinematic viscosity  $\nu$, are considered. More    technical details  are described  in the supplementary material. 
 
 \subsection{Colliding vortex rings}

\begin{figure}
\centering
\includegraphics[width=.9\linewidth]{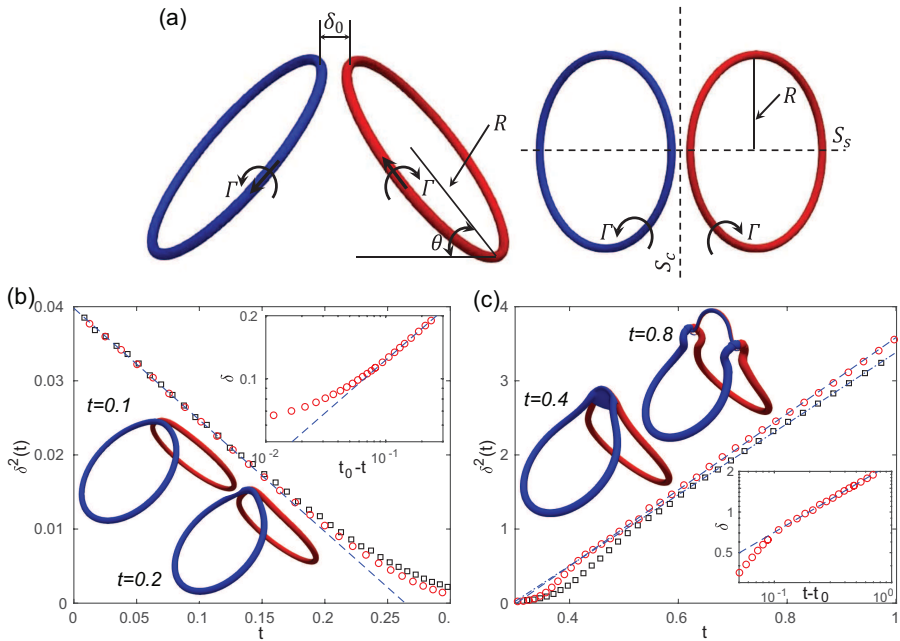}
\caption{Reconnection of  colliding vortex rings:  evolution of $\delta^2(t)$ at $Re_\Gamma=2000$ (\protect\marksymbol{square}{black}) and $4000$ (\protect\marksymbol{o}{red}) for (a) pre- and (b) post-reconnection phases.
The  blue dashed lines indicate the linear scaling. 
The insets  are  flow structures  represented by vorticity isosurface at 5\% of maximum initial vorticity $|\boldsymbol{\omega}| = 0.05\omega_0$ for $Re_\Gamma=2000$; and $\delta$ as a function of $|t-t_0|$ for $Re_\Gamma=4000$ with the dashed line referring the $t^{1/2}$ scaling. }
\label{fig:VRC_A45}
\end{figure}

We first consider the interaction of two  circular vortex rings, which  are symmetrically placed with the initial  inclination angle $\theta=\pi/4$  (figure \ref{fig:VRC_A45}a).
The initial radius of the ring is selected as $R_0=1$.
In addition, the initial minimum distance between these two vortex rings is chosen as $\delta_0=0.2$ so that the interaction between the vortices can be considered as   localized ($\sigma_0\ll \delta_0\ll R_0$). 
Note that this vortex setup  represents the typical antiparallel configuration. 
The evolution of the flow structure  for $Re_\Gamma=2000$ is shown in the insets of figure \ref{fig:VRC_A45} and also in supplementary movie S1. 
The structures for $Re_\Gamma= 4000$, which are quite similar, are not shown due to high computational cost for rendering.
Several features that distinctly differ from  quantum reconnection deserve to be noted. 
First, as the  rings approach each other under self-induction, they also undergo significant core deformation and form two thin vortex sheets.
Second, the reconnection process is not discrete as for quantized vortices; and circulation transfer rate and the reconnection time strongly depend on viscosity $\nu$, hence on  $Re_\Gamma$ \citep{hussain2011mechanics,yao2020physical}.
Finally, reconnection is never complete;  as a consequence, the circulation in the reconnected bridges is relatively smaller than the initial circulation $\Gamma_0$ of the vortices. 

The  appropriate determination of  $\delta(t)$ relies heavily on the accurate tracking of the location of the vortex axis \citep{fonda2014direct,villois2016vortex}, which is rather challenging in classical fluids.
First, unlike  vortex filaments or quantized vortices, where the axis location is almost precise, the vorticity field in classical fluids is continuously distributed. 
Second, vortex cores are typically distributed in irregular shapes without any clear center:  before reconnection, the vortices
undergo significant core deformation; and after reconnection, the reconnected vortex lines take some time to collect together to form the bridge.

Due to the  two-fold symmetry of the initial condition considered, the minimum distance $\delta$ between these two interacting rings before and after reconnection should occur in the symmetry $S_s$ and collision $S_c$ planes,  respectively -- which makes the determination of $\delta(t)$ relatively easy.  
Following \cite{hussain2011mechanics} and \cite{yao2020singularity},  we take the vorticity centroid (computed as the centroid of   above 75\% of its maximum) to be the center for vortices in these two planes.
Figure \ref{fig:VRC_A45}(a) displays the evolution of $\delta^2(t)$ for the pre-reconnection event, with the top inset showing $\delta$ as a function of $t_0-t$ on a log-log scale for $Re_\Gamma=4000$. 
The clear following of linear scaling for $\delta^2(t)$ at the early time  suggests that 
$\delta(t)\sim a^{-} (t^-_0-t)^{1/2}$,
with $a^{-}$ the constant prefactors for pre-reconnection corresponding to $A^-\Gamma^{1/2}$ in equation (\ref{eqn:dsca}), and $t^-_0$ the critical time when  $\delta \to0$.
For both $Re_\Gamma$  cases,
$\delta^2(t)$ collapses initially and then  slowly deviates from linear scaling when $\delta\sim\mathcal{O}(\sigma)$. 
The deviation happens earlier for the $Re_\Gamma=2000$  case,  which is due to the more rapid increase of the core size caused by stronger viscous diffusion. 
A linear fit on $\delta^2(t)$ between $0<t<0.15$ for $Re_\Gamma=4000$ gives $t^-_0=0.26$ and  $a^{-}=0.38$. 
As the circulation  remains constant at $\Gamma=1$ during this time, the dimensionless prefactor $A^{-}=a^{-}=0.38$, which is quite close to  $A=0.4$ reported in  \cite{PhysRevLett.72.482}.

When  two bridges move sufficiently apart from the interacting region, a clear linear scaling for $\delta^2(t)$ can be observed  for both $Re_\Gamma$ cases (figure \ref{fig:VRC_A45}b). 
Hence, $\delta\sim a^{+}(t-t^{+}_0)^{1/2}$ scaling also holds in the post-reconnection dynamics when the two bridges' vortices are mainly governed by the mutual interaction. 
The early evolution of $\delta^2(t)$ deviates from the linear scaling, presumably for two main reasons.
First, when the bridges are  too close,  they are under the influence of other unreconnected structures, such as threads, and other parameters besides $\Gamma$ may be relevant in determining $\delta$.  Second, the reconnected vortex lines, initially in a thin vortex sheet shape, take time to accumulate to form a circular shape, and
the circulation $\Gamma$ continuously increases during this phase.
A fit in the linear region gives $t^+_0\approx0.30$ for both $Re_\Gamma$ cases, and $a^+=2.19$ and $2.27$ for $Re_\Gamma=2000$  and  $4000$, respectively.
Different from quantized vortices, where reconnection is discrete and  $t_0$ is almost the same for pre- and post-reconnection, here reconnection is a continuous process, and hence $t^+_0$ is slightly larger than $t^-_0$.
Consistent with  previous studies, $a^+$ is always larger than $a^-$, indicating that the vortices separate much faster  than their approach.
Compared to the pre-reconnection process, the  effect of  $Re_\Gamma$ on $\delta(t)$ is more apparent for the post-reconnection. 
It is because, in classical fluids, the dynamics of reconnection, such as  the reconnection time and the  circulation transfer rate,  strongly depends on the viscosity $\nu$. 
In general, reconnection is faster at higher $Re_\Gamma$, which  explains why $\delta^2(t)$ follows linear scaling earlier at $Re_\Gamma=4000$.
In addition, as $Re_\Gamma$ increases,  reconnection is more complete \citep{yao2020singularity}.
The variation of $a^+$ with respect to $Re_\Gamma$ is mainly  attributed to different circulations $\Gamma$ in the reconnected bridges -- which is  difficult to be precisely determined.  

\begin{table}
\centering

\begin{tabular}{lrrrrr}
Cases  & $Re_\Gamma$&$a^-$ &$t^-_0$ &$a^+$ &$t^+_0$ \\
\multirow{2}{*}{1. Colliding vortex rings}&2000  & 0.38 & 0.26&2.19&0.30 \\
 &4000& 0.38 & 0.26&2.27&0.30 \\
\multirow{2}{*}{2. Orthogonal vortex tubes}&2000& 0.29 &0.60&0.94 &0.65\\
&4000& 0.29 &0.61&0.99 &0.63\\
\multirow{2}{*}{3. Vortex ring and tube}  &2000& 0.39 & 0.96&1.28&1.09\\
  &4000& 0.40 & 0.93&1.34&1.03\\
\end{tabular}
\label{table:sc}
\caption{Fitted values of the prefactors $a^{\pm}$ and $t^{\pm}_0$ for the minimum distance scaling $\delta(t)\sim a^{\pm}|t-t^\pm_0|^{1/2}$.
The superscript $\pm$ stands for  before ($-$) and after ($+$) the reconnection, respectively.}
\end{table}

 \subsection{Orthogonal vortex tubes}

\begin{figure}
\centering
\includegraphics[width=.95\linewidth]{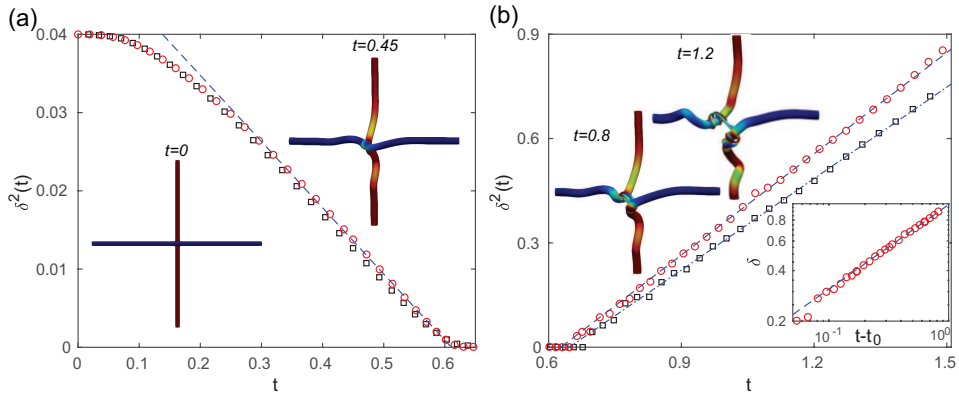}
\caption{Reconnection of orthogonal vortex tubes: time evolution of $\delta^2(t)$ at  $Re_\Gamma=2000$ (\protect\marksymbol{square}{black}) and $4000$ (\protect\marksymbol{o}{red}) for  (a) the pre- and (b) post-reconnection phases,  with the   dashed lines indicating  linear scaling. The insets are  flow structures represented by vorticity isosurface  $|\boldsymbol{\omega}| = 0.05\omega_0$;  the bottom inset in (b) is  $\delta$ as a function of $|t-t_0|$ for $Re_\Gamma=4000$ with the dashed line indicating the $t^{1/2}$ scaling. }
\label{fig:ON}
\end{figure}

As one of the simplest configurations, the reconnection of orthogonal vortex tubes  has been extensively  studied for both classical \citep{boratav1992reconnection,beardsell2016investigation,jaque2017reconnection} and quantum \citep{zuccher2012quantum,Galantucci12204} fluids.
 Similar to Case I, here the initial distance between these two rectilinear vortices is chosen as $\delta_0=0.2$
The insets in figures \ref{fig:ON}(a) and (b) and also  supplementary movie S3   show the evolution of the flow structures for $Re_\Gamma=2000$.
The evolution is quite similar to that in  \cite{boratav1992reconnection} for the thick vortex core case: the vortex tubes first develop into  locally antiparallel configuration under mutual induction; then collide with each other due to self-induction;  after reconnection, they recede away. 
Different from quantum cases \citep{villois2017universal,Galantucci12204}, the unreconnected threads, which wrap around the bridges,  are distinct after reconnection.
\textcolor{black}{In addition, a Kelvin wave is observed after reconnection.  In quantum fluids,  nonlinear interaction of Kelvin waves creates waves of shorter and shorter wavelength, which is considered as the main mechanism for energy cascade \citep{baggaley2011spectrum}; in classical fluids, however, the Kelvin wave would rapidly decay due to viscous effect.
It would be interesting to  compare the difference in the Kelvin wave evolution as well as its role on energy cascade between the quantum and classical  reconnections. }

\begin{figure}
\centering
\includegraphics[width=0.95\textwidth]{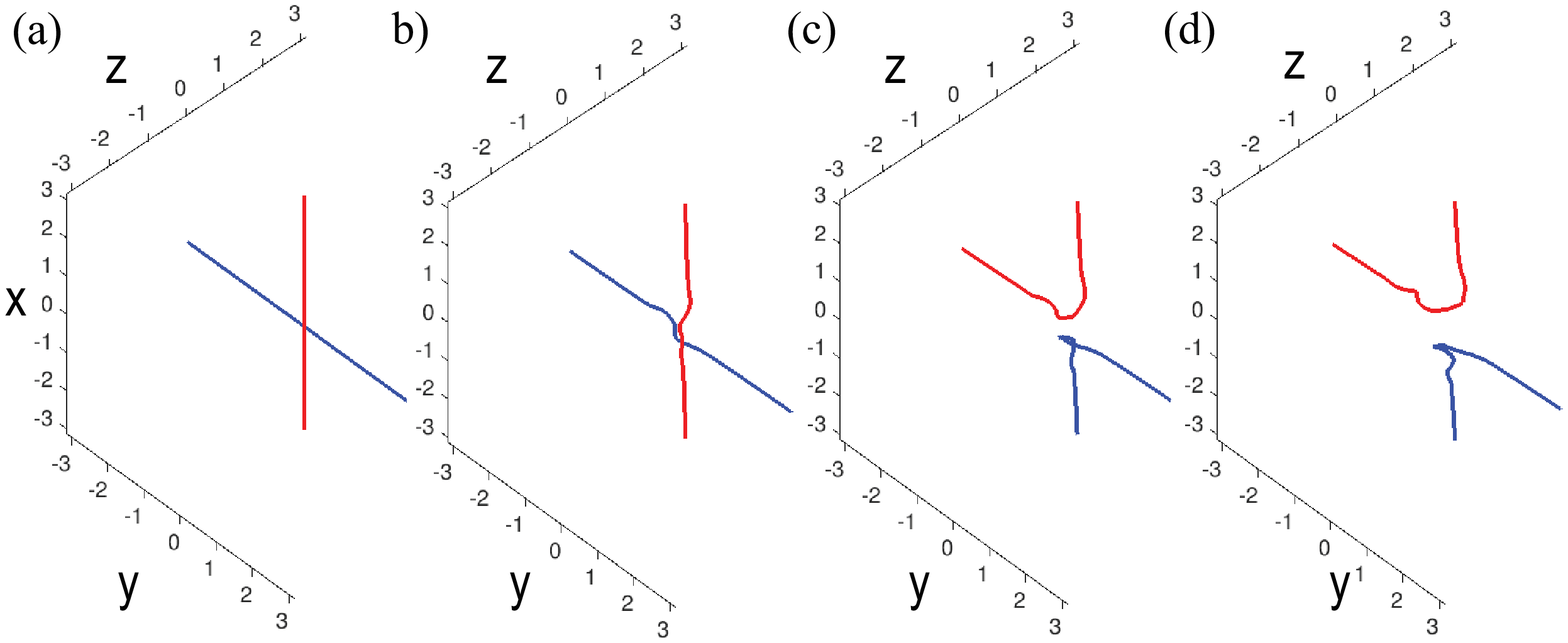}
\caption{Evolution of the  vortex axes for the orthogonal vortex tubes case at  $Re_\Gamma=2000$: (a) $t=0$, (b) $t=0.45$, (c) $t=0.8$; and $t=1.2$.}
\label{fig:VRC_ON}
\end{figure}

To determine the minimum distance $\delta(t)$ between these two vortex tubes,   the axis of the vortex tubes  needs to be tracked. 
Here, we propose a vortex tracking method based on the vortex lines that go through the vortex center  at the boundary. 
First, the  centriod of the vortex tubes at the planes $x=-\pi$ and  $y=-\pi$ is determined using same procedure as discussed above.
Then, vortex lines that seeds from these two centers are integrated using the ``stream3" function in Matlab.  
Figure \ref{fig:VRC_ON} (and  supplementary movie S4) shows the time evolution of the vortex axis for  $Re_\Gamma=2000$; and the evolution at $Re_\Gamma=4000$ are quantitatively the same.
\textcolor{black}{It is clear that the axis of vortex  tubes  is unambiguously identified.}
Finally,  $\delta$ is taken as the shortest distance between these two vortex lines. 


For the pre-reconnection, $\delta^2(t)$  initially varies slowly during the phase of the formation of anti-parallel configuration (figures \ref{fig:ON}a).
Then, the perturbed vortex tubes approach each other rapidly with  $\delta^2(t)$ follows a clear linear scaling. 
A slight difference in the evolution of $\delta^2(t)$ can be observed between $Re_\Gamma=2000$ and $4000$ cases, indicating a weak  Reynolds number effect on the pre-reconnection evolution.
For both $Re_\Gamma$ cases, the linear fit shows that $a^-\approx0.29$, which is slightly smaller than the case of the colliding rings.
Figure \ref{fig:ON}(b) shows that  $\delta^2(t)$  also follows linear scaling after reconnection, and the $\delta \sim t^{1/2}$ scaling extend far beyond the initial separation distance $\delta_0$.
Consistent with Case I, the prefactor increases with $Re_\Gamma$, with $a^+=0.93$ and $0.99$ for $Re_\Gamma=2000$ and $4000$, respectively.
Again, the vortices move faster after the reconnection than before it. 
Similar to the finding in  \cite{villois2017universal}, the prefactors $a^+$ are  smaller than those in Case I, which might be due to a smaller curvature of the cusps generated after reconnection in this case.

 \subsection{Vortex ring and tube interaction}

\begin{figure*}
\centering
\includegraphics[width=.9\linewidth]{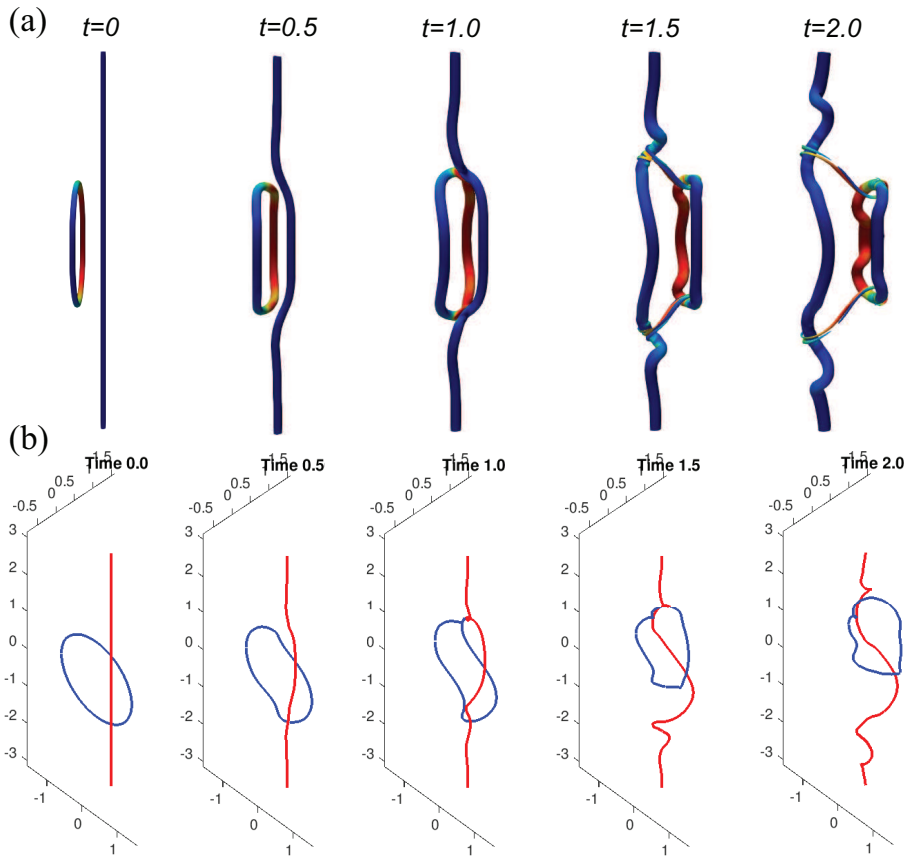}
\caption{\textcolor{black}{Evolution of flow structures  for   vortex ring and tube interaction for $Re_\Gamma=2000$: (a)  represented by vorticity isosurface at 5\% of maximum initial vorticity, i.e. at $|\boldsymbol{\omega}| = 0.05\omega_0$; and (b) by tracked vortex axis. } }
\label{fig:RT1}
\end{figure*}

The third case we considered is a vortex ring interacting with an isolated  rectilinear vortex tube. 
The radius of the ring is chosen the same as the colliding vortex rings case, namely, $R_0=1$. 
\textcolor{black}{To reveal the crossover from driven ($\delta\sim t$) to interaction ($\delta\sim t^{1/2}$) region observed in  \cite{Galantucci12204}, the initial distance is chosen as twice  the previous cases, namely,  $\delta_0=0.4$. }
The vortex setup and the subsequent evolution for $Re_\Gamma=2000$ represented by vortex surfaces and tracked vortex axis are shown in the top insets in  figure \ref{fig:RT1}(a) and (b), respectively (see also supplementary movies S5 and S6). 
Due to the self-induction, the vortex ring approaches the vortex tube; during this phase, both the vortex ring and tube are perturbed; at close approach, the vortex ring and tube are also deformed into locally antiparallel configuration (i.e., $t=1$). 
It further confirms the argument that reconnection physics of two vortices should be independent of the initial spatial configuration \citep{siggia1985incipient}.
After  reconnection, parts of the vortex ring and tube exchange with each other; and due to the Kelvin wave, the newly formed vortex ring and tube  become further perturbed with  the threads connecting  them. 

\begin{figure*}
\centering
\includegraphics[width=.95\linewidth]{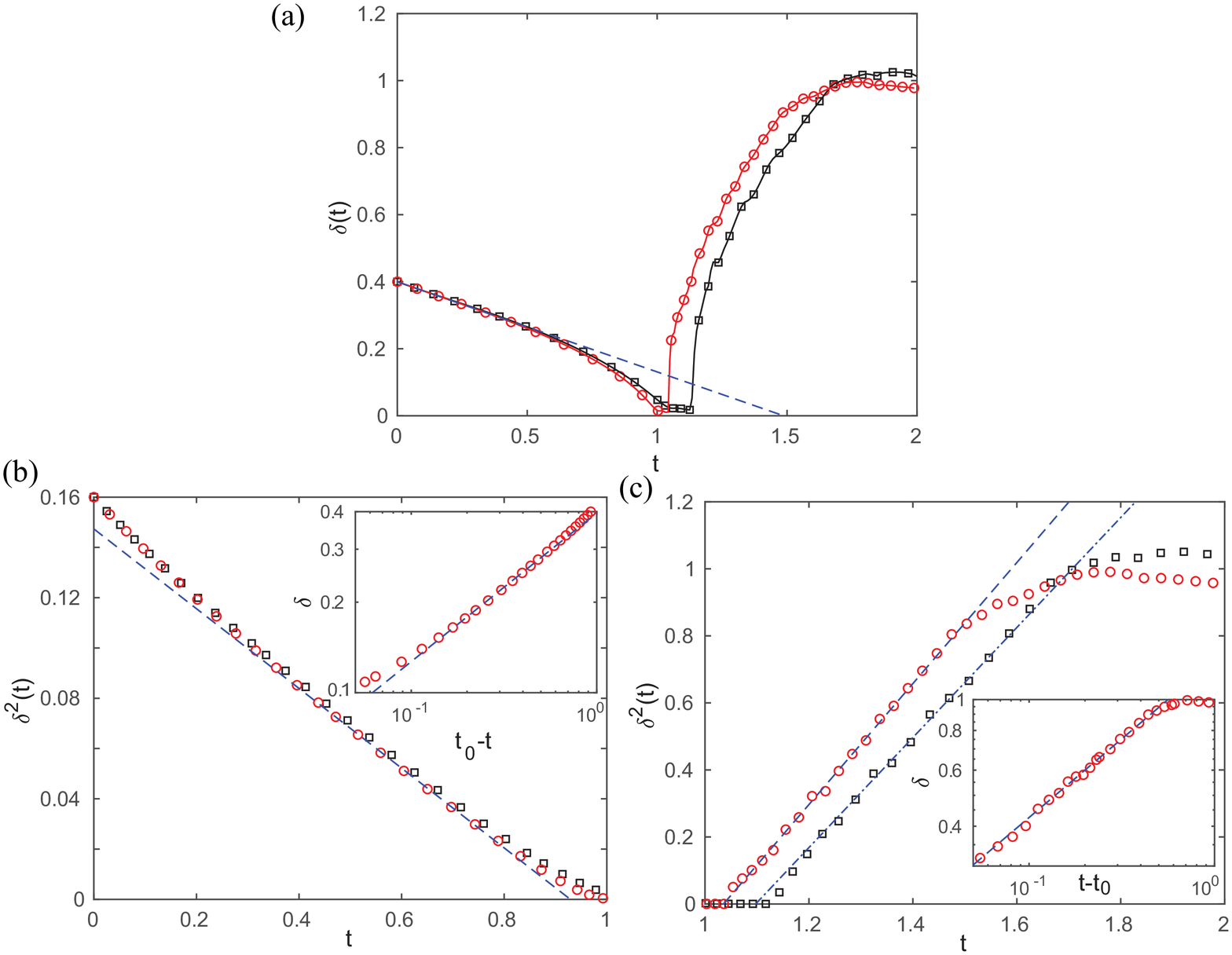}
\caption{\textcolor{black}{Interaction of vortex ring and tube: (a) 
time evolution of $\delta(t)$; and $\delta^2(t)$ for  (b) the pre-reconnection and (c) post-reconnection phases. 
 Symbols \protect\marksymbol{square}{black} and \protect\marksymbol{o}{red} refer to $Re_\Gamma=2000$ and $4000$, respectively; and the  blue dashed lines indicate linear scaling. The insets  in (b) and (c) show separation distance $\delta$ as a function of $|t-t_0|$ for $Re_\Gamma=4000$ with the dashed line indicating the $t^{1/2}$ scaling. }}
\label{fig:RT2}
\end{figure*}

Figure \ref{fig:RT2}(a) displays the evolution of  $\delta(t)$ with figure \ref{fig:RT2}(b) and (c) showing  $\delta^2(t)$ before and after reconnection, respectively.
Initially, $\delta(t)$  scale almost linearly with $t$ and the approaching velocity can be approximately determined by the initial self-induced velocity of the ring and the mutual-induced velocity between the ring and the tube.  
Consistent with the previous two cases, when the two vortices are close to each other, a clear $t^{1/2}$ scaling for $\delta$ is observed (inset in figure \ref{fig:RT2}b). 
The transition between driven ($\delta\sim t$) and interaction ($\delta\sim t^{1/2}$) regions   happens at $\delta\sim0.3$.
The prefactor for $Re_\Gamma=4000$ is $a^-=0.40$, which is very close to Case I.

From  figure \ref{fig:RT2}(c), it is  clear that $\delta(t)\sim t^{1/2}$ scaling holds after reconnection, with the prefactor $a^+=1.28$ and $1.34$ for $Re_\Gamma=2000$ and $4000$, respectively.  
The values are between the colliding vortex rings and orthogonal tubes cases.
The $1/2$ scaling breaks down when the vortex ring moves sufficiently far away from the tube. Note that the crossover between the $t^{1/2}$ to $t^1$ scalings for $\delta(t)$ in the post-reconnection is not observed.
Instead,  for this case $\delta(t)$ remains almost constant at late times. 
The reason is  that the traveling velocity of the perturbed vortex ring is roughly the same as that of the perturbed part of the tube.
When the oscillations in the vortex tube and ring die out and the vortex ring regains its circular shape, we should expect $\delta\sim t$ as suggested in  \cite{Galantucci12204}.

 \section{Conclusions}

The question of whether there is a universal scaling/route for  reconnection has been extensively studied and debated \citep{zuccher2012quantum,villois2017universal,fonda2019reconnection}. 
Prior works on quantum vortex reconnection have shown clear evidence for the existence of a  universal $\delta \sim t^{1/2}$ scaling; however, due to the complex nature  for reconnection in classical fluids (presumably due to  viscosity), this scaling has never been confirmed previously. 
With the aid of recent advances in supercomputing, we performed   direct numerical simulation of  viscous reconnection for slender vortices at $Re_\Gamma=2000$ and $4000$.
Three different initial conditions are considered, namely, two colliding vortex rings; orthogonal and straight vortex tubes; and vortex ring interacting with a tube. 
For all these cases, the vortices evolve into locally antiparallel configuration -- akin to the finding in  \cite{villois2017universal} for the reconnection of quantum vortices.
When the distance between two interacting vortices is large compared with their core size, and the dynamics are predominately governed by their mutual induction,  we observe, for the first time, that the approach and separation distances follow a symmetrical 1/2-power scaling, independent of the initial  configuration.
The discrepancies in previous studies \citep{hussain2011mechanics,yao2020physical} are  due to the fact that the  length scale of vortex core size $\sigma$ is approximately the same order as the separation $\delta$ and  should  be incorporated when considering the scaling.  
Although the dynamics of the reconnection is substantially different from that in quantum fluids, the surprisingly similar results in classical fluids regarding $\delta(t)$  scaling   suggest that there is indeed a universal route towards reconnection. 
Consistent with previous results \citep{zuccher2012quantum,boue2013analytic, villois2017universal}, we find that the prefactors $a^{\pm}$ in the square root law is not universal and  depend on the initial configuration as well as the  Reynolds number (or viscosity) -- which is a distinct feature for classical  vortex reconnection.

\section*{Acknowledgments}
Computational  resources provided by Texas Tech University HPCC, TACC Lonestar and Frontera are acknowledged; and visualization using XSEDE Stampede2 is also appreciated.  
The original data for the separation distance scaling presented in figures \ref{fig:VRC_A45}-\ref{fig:RT2} can be downloaded from   Texas Data Repository Dataverse
https://doi.org/10.18738/T8/ONA8DG; and 
the full flow field data are available   from   the   authors  upon reasonable request.

\section*{Declaration of interests}
The  authors  report  no  conflict  of  interest.

\appendix

\bibliographystyle{jfm}
\bibliography{fluidmechanics}
\end{document}